\documentclass[
    ,final            % use final for the camera ready runs
  ]
  {aipproc}

\layoutstyle{8x11double}

\usepackage{graphicx,natbib}
\usepackage{amssymb}
\newcommand{\be}{\begin{equation}}
\newcommand{\ba}{\begin{eqnarray}}
\newcommand{\ee}{\end{equation}}
\newcommand{\ea}{\end{eqnarray}}

\def\lesssim{\mathrel{\hbox{\rlap{\hbox{\lower4pt\hbox{$\sim$}}}\hbox{$<$}}}}
\def\gtrsim{\mathrel{\hbox{\rlap{\hbox{\lower4pt\hbox{$\sim$}}}\hbox{$>$}}}}

\def\simless{\mathbin{\lower 3pt\hbox
   {$\rlap{\raise 5pt\hbox{$\char'074$}}\mathchar''7218$}}}   % < or of order
\def\simgreat{\mathbin{\lower 3pt\hbox
   {$\rlap{\raise 5pt\hbox{$\char'076$}}\mathchar''7218$}}}   % > or of order
\def\apj{ApJ}
\def\apjs{ApJS}
\def\apjl{ApJL}

\def\mnras{MNRAS}

\def\physrep{{Phys. Reports}}
\def\prd{{Phys. Rev. D}}
\begin{document}

\title{Simulating Reionization: Character and Observability}

%<Replace this text with PACS numbers; choose from this list:
%                \texttt{http://www.aip..org/pacs/index.html}>}

\author{Ilian T. Iliev}{
  address={Canadian Institute for Theoretical Astrophysics, University
  of Toronto, 60 St. George Street, Toronto, ON M5S 3H8, Canada}
}

\author{Paul R. Shapiro}{
  address={Department of Astronomy, University of Texas, Austin, TX 78712-1083,
  U.S.A.}
}

\author{Garrelt Mellema}{
  address={Stockholm Observatory, AlbaNova
  University Center, Stockholm University, SE-106 91 Stockholm, Sweden}
%<common address for author2 and author3>}
%  ,altaddress={<author1 address>} % additional visiting address
}

\author{Ue-Li Pen}{
  address={Canadian Institute for Theoretical Astrophysics, University
  of Toronto, 60 St. George Street, Toronto, ON M5S 3H8, Canada}
}
\author{Patrick McDonald}{
  address={Canadian Institute for Theoretical Astrophysics, University
  of Toronto, 60 St. George Street, Toronto, ON M5S 3H8, Canada}
}
\author{J. Richard Bond}{
  address={Canadian Institute for Theoretical Astrophysics, University
  of Toronto, 60 St. George Street, Toronto, ON M5S 3H8, Canada}
}

\begin{abstract}
In recent years there has been considerable progress in our 
understanding of the nature and properties of the reionization 
process. In particular, the numerical simulations of this epoch
have made a qualitative leap forward, reaching sufficiently large
scales to derive the characteristic scales of the reionization 
process and thus allowing for realistic observational predictions. 
Our group has recently performed the first such large-scale 
radiative transfer simulations of reionization, run on top of 
state-of-the-art simulations of early structure formation. This 
allowed us to make the first realistic observational predictions 
about the Epoch of Reionization based on detailed radiative 
transfer and structure formation simulations. We discuss the basic 
features of reionization derived from our simulations and some 
recent results on the observational implications for the 
high-redshift Ly-$\alpha$ sources.
\end{abstract}

\maketitle

%%%%%%%%%%%%%%%%%%%%%%%%%%%%%%%%%%%%%%%%%%%%
%% MAINMATTER
%%%%%%%%%%%%%%%%%%%%%%%%%%%%%%%%%%%%%%%%%%%%

\section{Introduction}   %%% Top level section head (remove ``%'' symbol)
The reionization of the universe was the last global transition of 
the Intergalactic Medium (IGM), caused by the radiation of the First Stars.
It completely transformed the IGM, from fully-neutral and cold to almost 
fully-ionized and hot ($T_{\rm IGM}\sim10^4$~K). Reionization started with 
the formation of the first sources of ionizing radiation at very high
redshifts $z\sim30-40$ and finished around $z\sim6-7$. Currently there are 
still very few direct observational constraints on this epoch. The lack of 
Gunn-Peterson trough in the spectra of high-redshift sources implies low 
mean neutral fraction $x_{\rm HI}\lesssim10^{-4}$ out to redshift $z\sim5.5$,
which appears to rise at higher redshifts,
possibly indicating that at these redshifts we are seeing the tail-end of 
reionization. The second direct observable, the integrated 
electron-scattering optical depth, $\tau_{\rm es}$, was found to be 
$0.09\pm0.03$ based on the WMAP 3-year data \citep{2007ApJS..170..377S}. 
This range of values 
points to an early start and extended period of reionization. However, 
the error bars remain fairly large and thus the optical depth by itself 
does not put very stringent constraints on reionization. 
%Furthermore, the Jeans-mass filtering of low-mass sources in 
%the ionized regions results in the $\tau_{\rm es}$ and $z_{\rm ov}$, the 
%overlap redshift, being only loosely related \citep{2007MNRAS.376..534I}. 
%The reason for this is that $z_{\rm ov}$ depends on the abundances and 
%efficiencies of the high-mass sources, whose formation is not suppressed 
%by reionization, while $\tau_{\rm es}$ itself depends on both types of 
%sources. Thus, varying the ionizing efficiencies of low-mass sources 
%yields a wide range of $\tau_{\rm es}$ values for the same $z_{\rm ov}$. 
%Therefore, much more data is needed to put any robust constraints on the 
%possible reionization scenarios and the properties of the high-redshift 
%galaxies.  

\begin{figure}
%\begin{center}
  \includegraphics[width=5.2in]{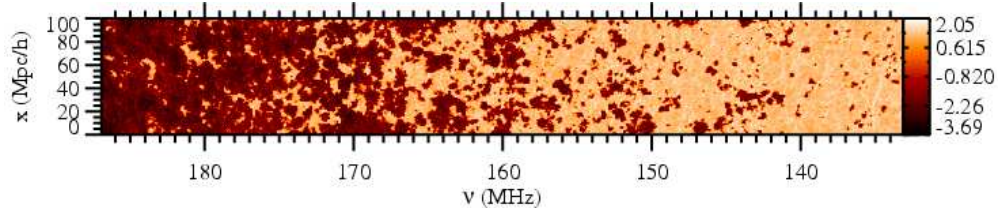}
%  \includegraphics[width=3.2in]{f250C_slice_beambw.ps}
%\vspace{-0.9cm}
\caption{Position-redshift slices from one of our simulations,
%the f250C simulation. These slices
  illustrating the large-scale geometry of reionization and the significant 
  local variations in reionization history as seen at the redshifted 21-cm 
line. 
%  Observationally they correspond to slices through an image-frequency
%  volume. 
Shown is the decimal log of the differential brightness temperature. 
% at the full grid resolution.
%  The spatial scale is in comoving units.
\label{pencil}}
%\end{center}
\end{figure}

\begin{figure}
  \includegraphics[width=1.5in]{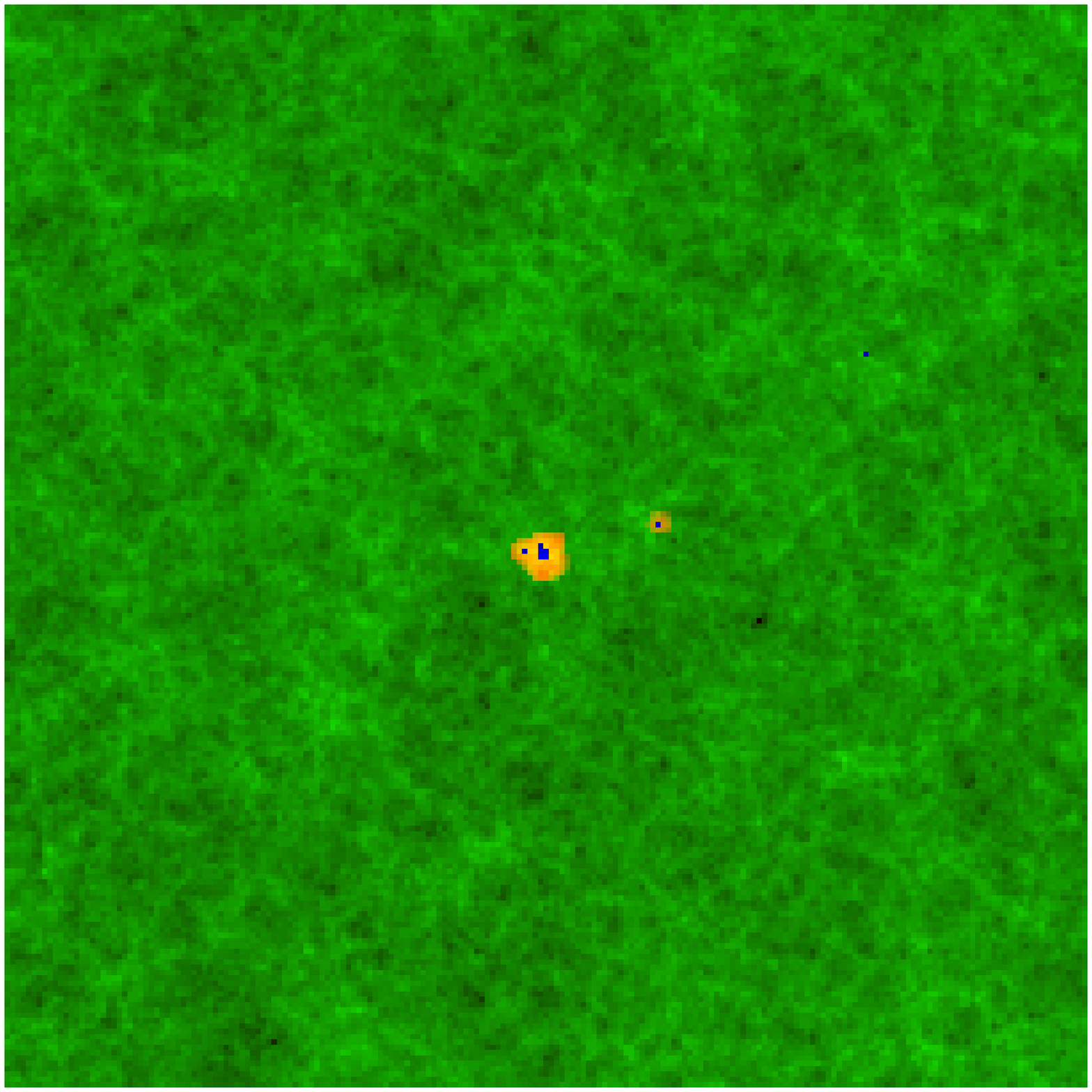}
  \includegraphics[width=1.5in]{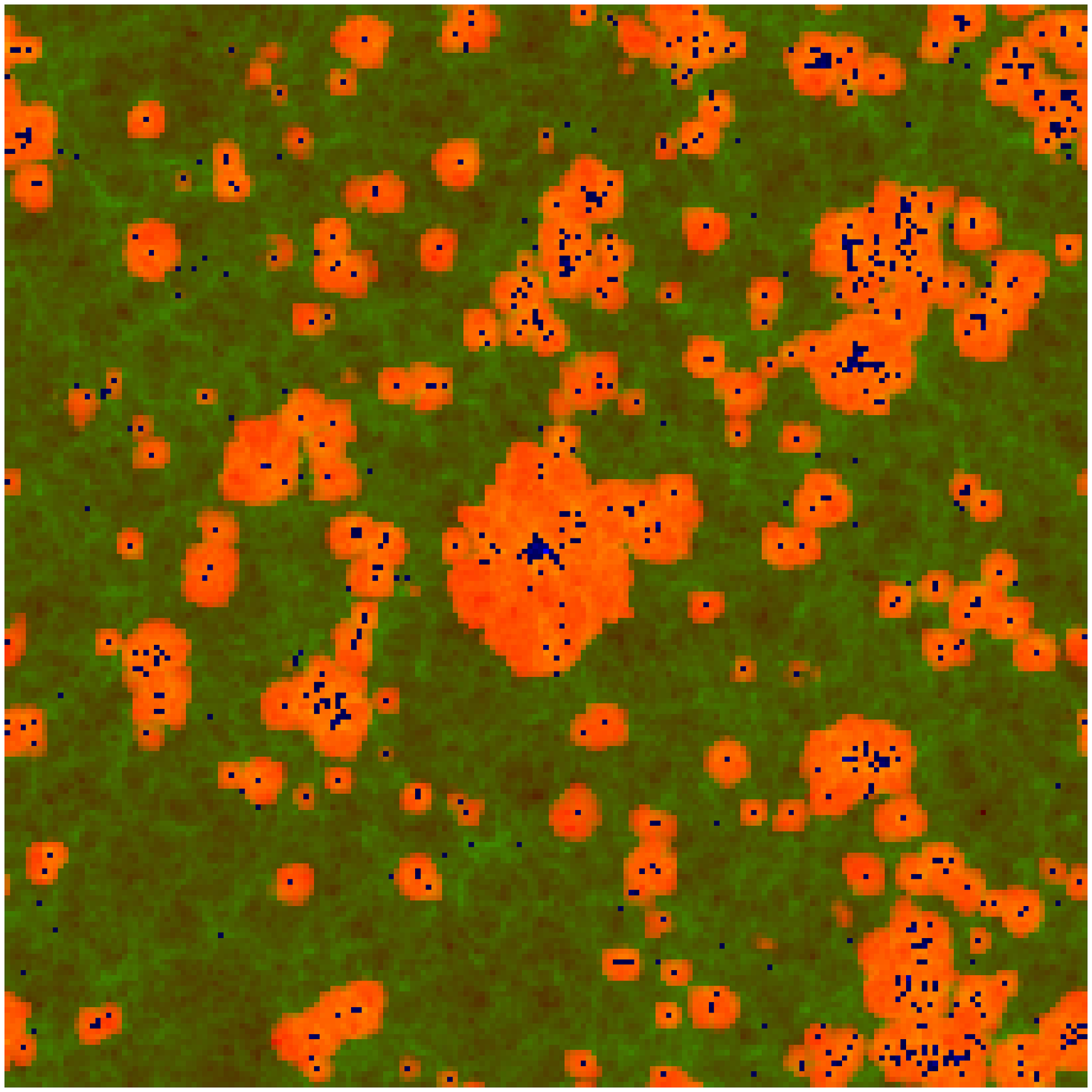}
  \includegraphics[width=1.5in]{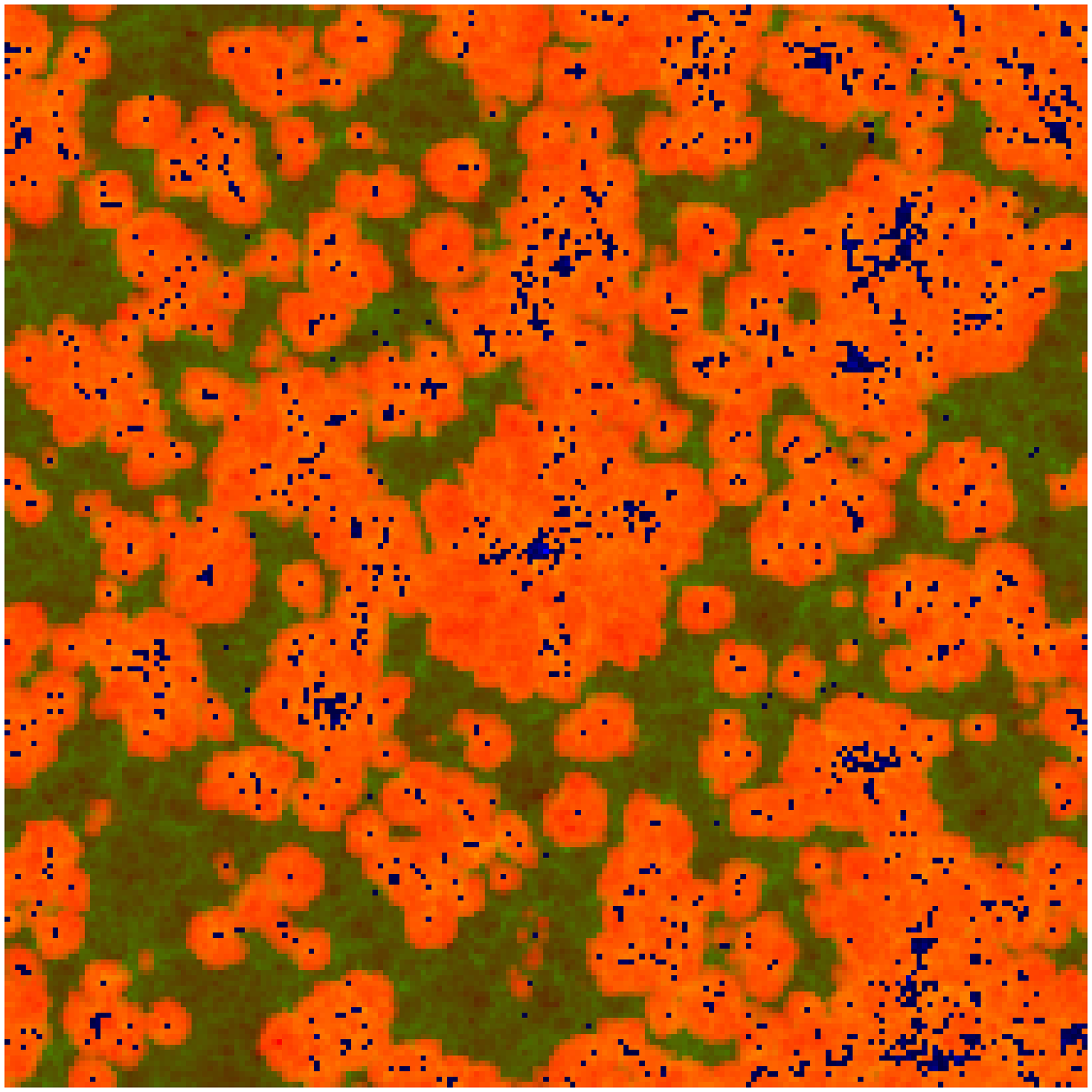}
  \includegraphics[width=1.5in]{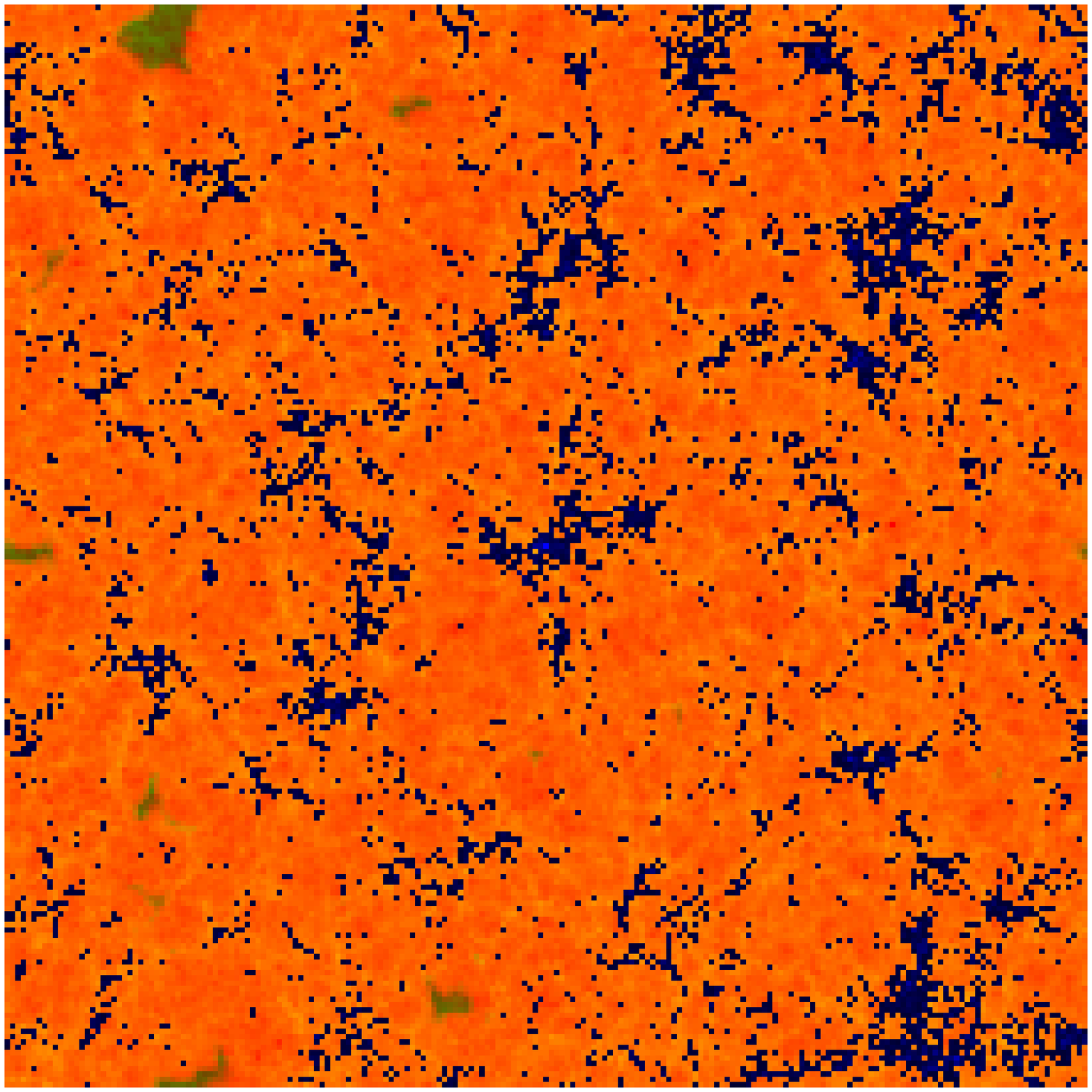}
\caption{The reionization history of a high density peak.
The images are centered on the most massive (at $z=6$) halo
in our computational volume and are of size $100\,h^{-1}$Mpc
to the side. The snapshots are (left to right): $z=12.9$,
$z=9.0$, $z=8.0$, and $z=6.6$. The underlying cosmological 
density field (dark,green) is superimposed with the ionized 
fraction (light, orange) and the ionizing sources (dark, blue 
dots).
\label{peak_evol}}
\end{figure}

This relative lack of observational data is set to change dramatically in 
the next few years, however. A number of large observational projects are 
currently under development, e.g. observations at the redshifted 21-cm 
line of hydrogen \citep[e.g.][]{1997ApJ...475..429M,2000ApJ...528..597T,
2002ApJ...572L.123I,2006MNRAS.372..679M,2006PhR...433..181F}, detection of 
small-scale CMB anisotropies due to the kinetic Sunyaev-Zel'dovich (kSZ) 
effect \citep[e.g.][]{2000ApJ...529...12H,2003ApJ...598..756S,kSZ}, and 
surveys of high-redshift Ly-$\alpha$ emitters 
\citep[e.g.][]{%2003AJ....125.1006R,
2004ApJ...604L..13S,2006NewAR..50...94B}.
The planning and success of these experiments relies critically upon 
understanding the large-scale topology of reionization, i.e. the size- and 
spatial distribution of the ionized and neutral patches, which is best 
derived by large-scale simulations. Recently we presented the first 
large-scale, high-resolution radiative transfer simulations of cosmic 
reionization
\citep{2006MNRAS.369.1625I,2006MNRAS.372..679M,2007MNRAS.376..534I}. Here 
we summarize our basic conclusions about the character of the 
reionization process, along with some of our recent results on its 
observability. %Throughout this work we assume a flat 
%($\Omega_k=0$) 
%$\Lambda$CDM cosmology %\\
%($\Omega_m,\Omega_\Lambda,\Omega_b,h,\sigma_8,n)=(0.24,0.76,0.042,0.73,0.74,
%0.95)$ 
%based on WMAP 3-year results \citep{2006astro.ph..3449S}, hereafter WMAP3. 
%Here $\Omega_m$, $\Omega_\Lambda$, and $\Omega_b$ are the total matter,
%vacuum, and baryonic densities in units of the critical density, $\sigma_8$ is
%the rms density fluctuations extrapolated to the present on the scale of 
%$8 h^{-1}{\rm Mpc}$ according to the linear perturbation theory, and $n$ is 
%the index of the primordial power spectrum of density fluctuations. 
%In this work we consider one specific run, labelled f250C \citep[for details 
%on the simulations and our notation see][]{2007MNRAS.376..534I}.

\section{Basic features of reionization}

The key results for the nature and features of reionization could be
summarized as follows:
\begin{itemize}
\item Reionization proceeds in an inside-out fashion, whereby the high 
density regions are ionized first and the deepest voids - last, due to 
the galaxies preferentially forming in and around the high density peaks 
of the density distribution \citep{2006MNRAS.369.1625I}.
\item The process is thus highly-inhomogeneous and patchy on large scales,
with very complex geometry. It is strongly modulated by the long-wavelength 
density fluctuations. As a consequence, small-volume simulations, which do 
not include these large-scale fluctuations underestimate the source biasing 
and the reionization patchiness
\citep{2006MNRAS.369.1625I,2006MNRAS.372..679M,wmap3}.
\item The reionization parameters, like source efficiencies (dependent on the 
stellar IMF, star formation efficiencies and escape fractions) and small-scale
gas clumping, are only weakly constrained at present. Both the overlap epoch,
$z_{\rm ov}$ and the integrated optical depth $\tau_{\rm es}$ are readily
reproduced by a wide range of reasonable reionization scenarios
\citep{2006MNRAS.369.1625I,2006MNRAS.372..679M,2007MNRAS.376..534I}.  
\item The H~II regions have typical scales of 5-20 comoving Mpc (dependent on
still-uncertain parameters). Although the high-redshift sources are typically 
very low-mass, their strong clustering results in the formation of such large 
ionized patches (see Figures~\ref{pencil} and \ref{peak_evol}). These
characteristic scales are directly reflected in many of the predicted
reionization observables \citep{2006MNRAS.372..679M,kSZ,pol21,cmbpol,wmap3}.
\item Reionization is strongly self-regulated through Jeans-mass filtering
of the low-mass sources: more efficient low-mass sources (e.g. due to
top-heavy IMF) result in more suppression
of the same sources, thereby largely cancelling the effect of the higher
efficiency, and vice versa \citep{2007MNRAS.376..534I}.  
\item This process of self-regulation has the effect that $\tau_{\rm es}$ 
and $z_{\rm ov}$, the overlap redshift are only loosely related and do not 
impose very strong constraints on the possible reionization scenarios. The
reason for this is that $z_{\rm ov}$ depends on the abundances and
efficiencies of the high-mass sources, whose formation is not suppressed by
reionization, unlike $\tau_{\rm es}$, which depends on both types of
sources. Varying the ionizing efficiencies of low-mass sources yields a wide
range of $\tau_{\rm es}$ values for the same $z_{\rm ov}$
\citep{2007MNRAS.376..534I}. 
\end{itemize}

\begin{figure*}
  \includegraphics[width=2.2in]{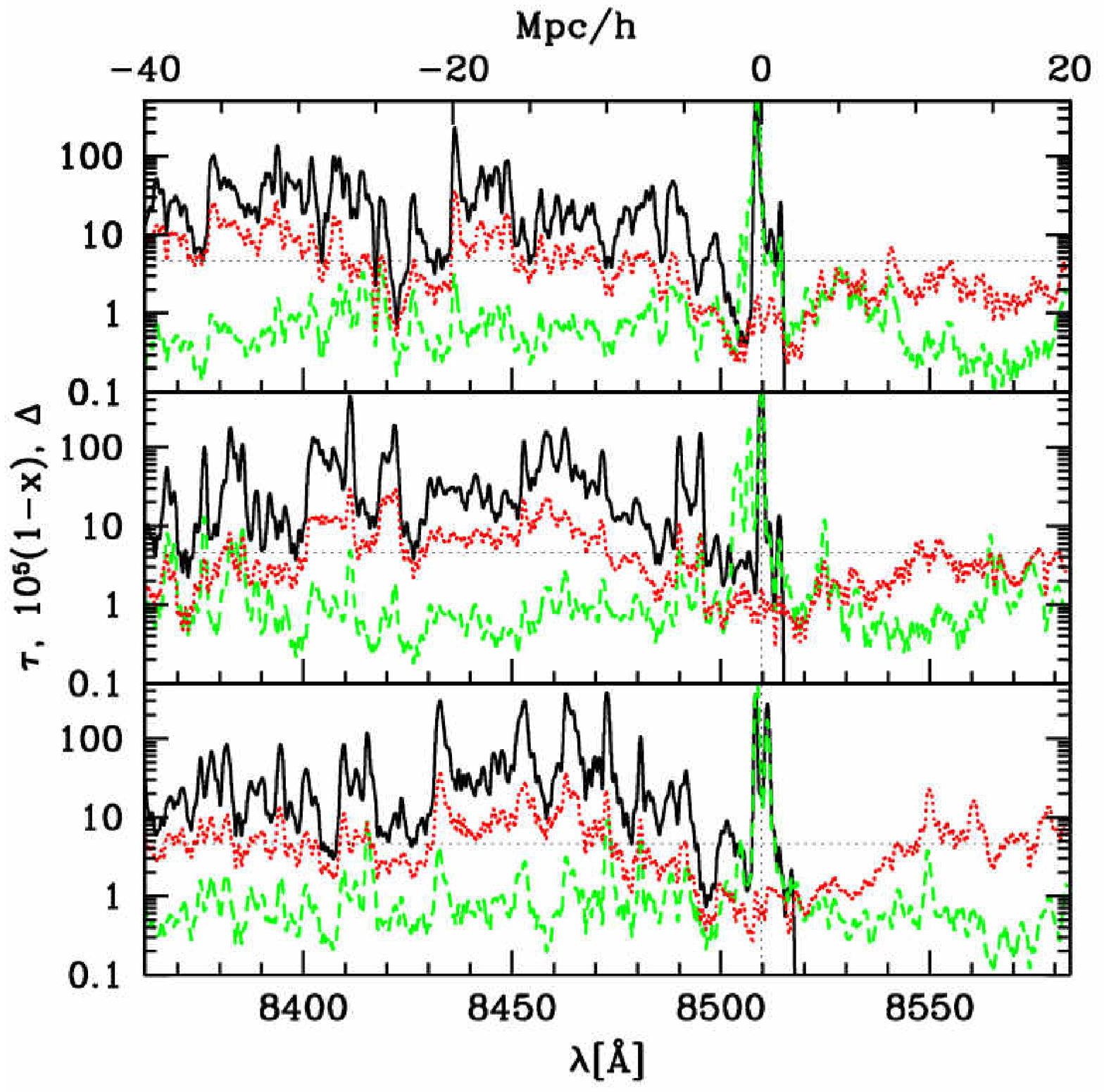}
  \includegraphics[width=2.2in]{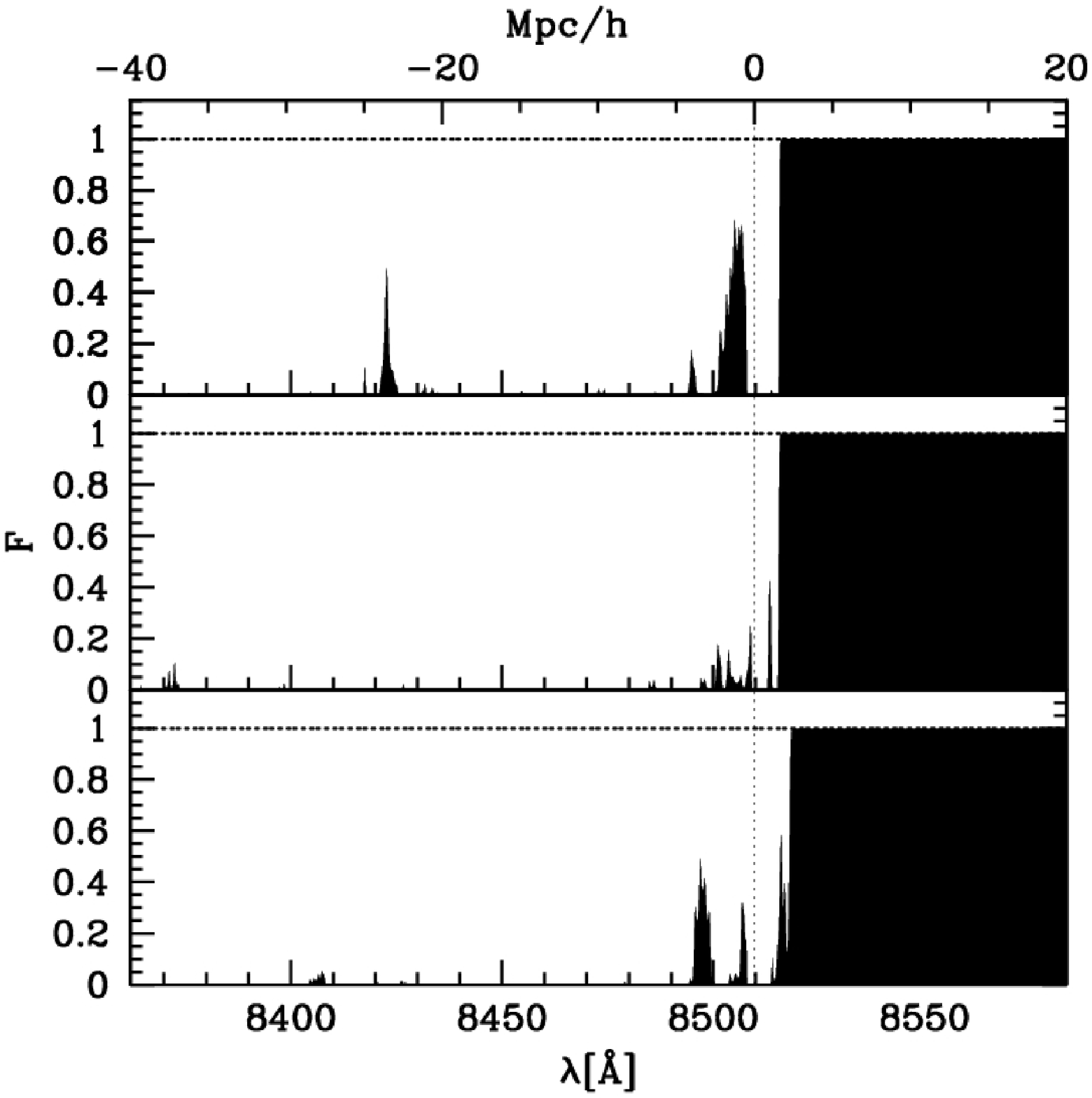}
  \includegraphics[width=2.2in]{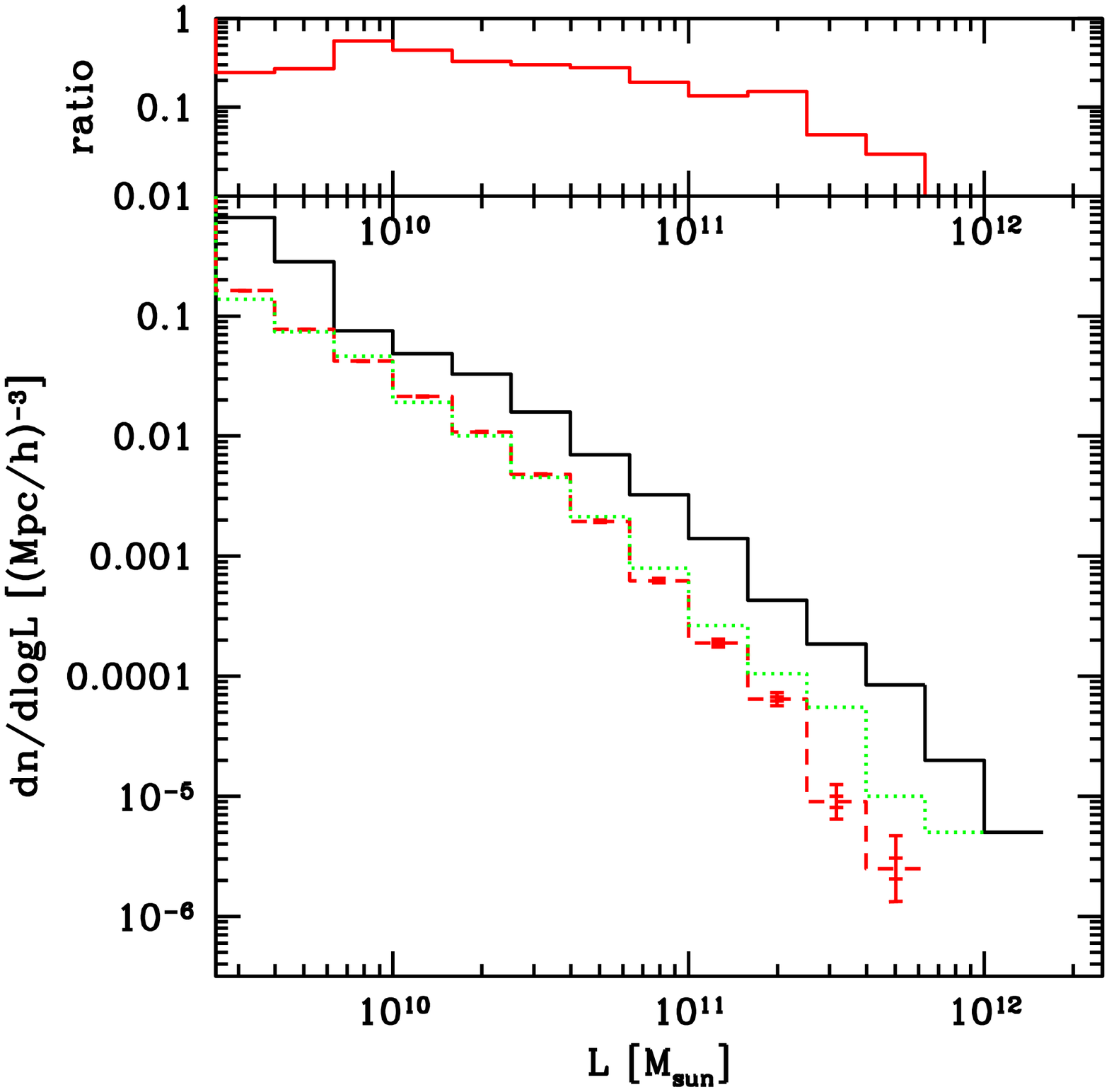}
\caption{Ly-$\alpha$ sources at redshift $z=6.0$: (left and middle) Sample LOS 
vs. $\lambda/$comoving distance from the most massive galaxy. Shown are (left)
the optical depth (solid), neutral fraction $\times10^4$ (dotted) and density
(dashed), and (right) the corresponding transmission. The vertical lines show 
the position of the central source (in redshift space, i.e. accounting for 
its peculiar velocity), while the horizontal lines indicate the optical depth
equivalent to 1\% transmission. Redshift-space distortions due to the local 
peculiar velocities and the Ly-$\alpha$ line damping wing are included; and 
(right) Luminosity function without 
(black) and with absorption included (red). For reference, the green, dotted 
line shows the result of suppressing each source by 50\%, which would be the case
if e.g. all of the blue wing of the emission line, but none of the red wing is
absorbed. 
\label{spectra}}
\end{figure*}

\section{High-redshift Ly-$\alpha$ sources}
Observations of the high-redshift Ly-$\alpha$ sources, galaxies and QSO's,
have already provided us with a wealth of information about the state of the
IGM and the nature of the luminous galaxies at the end of reionization and
still hold a lot of promise for the future. In addition to probing the IGM and
the source luminosity function (and thus, indirectly, the halo mass function
and ionization source properties), they may also be used to constrain the
reionization geometry. The first rare objects form in the highest peaks of the
density field. The statistics of Gaussian fields predicts that density peaks
are strongly clustered, especially at high redshift. As a consequence, each 
high-redshift, massive galaxy was surrounded by numerous smaller ionizing 
sources. Self-consistent simulations of such regions require following 
a large enough volume while at the same time resolving all the low-mass halos 
which drive reionization, which has not been possible until recently. 

In Figure~\ref{peak_evol} we illustrate several stages of the reionization 
history of a high density peak. The most massive source in our volume (at
$z=6$) is shifted to the centre using the periodicity of the computational 
box. At redshift $z=12.9$ the H~II region is small and the source is invisible
due to the damping wing due to the neutral gas outside. By redshift $z=9$ 
many more haloes have formed, most of them in large clustered groups. The 
H~II region surrounding the central peak is among the largest, but the central
source emission is still strongly affected by damping. Only by redshift $z=8$
is the ionized region large enough to render the source unaffected by damping
and potentially visible. The reionization geometry becomes quite complex, most 
ionized bubbles are interconnected, but large neutral patches remain between 
them. Finally, by the nominal overlap $z=6.6$ all ionized regions have merged 
into one topologically-connected region, although substantial neutral patches
remain interspersed throughout our volume, which remains largely
optically-thick to both Ly-$\alpha$ and ionizing continuum radiation. Only by 
$z\sim6$ this volume becomes on average optically-thin to ionizing radiation.

In Figure~\ref{spectra} (left and center) we show some sample spectra for the
same luminous source. The spectra exhibit extended high-transmission (10-60\%
transmission) regions in the highly-ionized proximity zone of the luminous
source, within 5~Mpc$\,h^{-1}$ ($\sim20$~\AA). The center of the peak itself
is optically-thick due to its high density. The infall around the central peak
blue-shifts photons, resulting in some absorption behind the redshift-space
position of the source. Away from the proximity region the absorption is
largely saturated, but there are a number of transmission gaps with up to a
few per cent transmission. In Figure~\ref{spectra} (right) we show the
Ly-$\alpha$ source luminosity function at the same redshift. For the weaker 
sources  roughly half of the intrinsic luminosity is transmitted (the red wing
of the line), while the most luminous sources suffer from additional
absorption due to the gas infall that surrounds them. Future work will
quantify the statistics of these features and its evolution. 

%\subsection{}   %%% Second level section head (remove ``%'' symbol)
%\subsubsection{}   %%% Lowest level section head (remove ``%'' symbol)
%\section*{}    %%% Unnumbered top level section head (remove ``%'' symbol)
%\subsection*{}   %%% Unnumbered second level section head (remove ``%'' symbol)

%%%%%%%%%%%%%%%%%%%%%%%%%%%%%%%%%%%%%%%%%%%%%%%%
%% BACKMATTER
%%%%%%%%%%%%%%%%%%%%%%%%%%%%%%%%%%%%%%%%%%%%%%%%

%\begin{theacknowledgments}
%This work was partially supported by NASA Astrophysical Theory Program grants
%NAG5-10825 and NNG04G177G to PRS.
%\end{theacknowledgments}

\bibliographystyle{aipproc}   % if natbib is available

%\bibliography{../../refs.bib}

\end{document}